\documentclass[twocolumn,showpacs,preprintnumbers,amsmath,amssymb,pre]{revtex4}

\usepackage{graphicx}
\usepackage{dcolumn}
\usepackage{bm}
\usepackage{enumerate}
\usepackage{amsmath}
\usepackage{color}

\begin{document}

\title{Stochastic foundations of $g$-subdiffusion process}

\author{Tadeusz Koszto{\l}owicz}
 \email{tadeusz.kosztolowicz@ujk.edu.pl}
 \affiliation{Institute of Physics, Jan Kochanowski University,\\
         Uniwersytecka 7, 25-406 Kielce, Poland}

 \author{Aldona Dutkiewicz}
 \email{szukala@amu.edu.pl}
 \affiliation{Faculty of Mathematics and Computer Science,\\
Adam Mickiewicz University, Uniwersytetu Pozna\'nskiego 4, 61-614 Pozna\'n, Poland}

\date{\today}

\begin{abstract}
Recently, in the paper: T. Koszto{\l}owicz and A. Dutkiewicz, Phys. Rev. E \textbf{104}, 014118 (2021) the $g$--subdiffusion equation with fractional Caputo time derivative with respect to another function $g$ has been considered. This equation offers new possibilities for modelling diffusion such as a process in which a type of diffusion evolves continuously over time. However, the equation has not been derived from a stochastic model and the stochastic interpretation of $g$--subdiffusion has been unknown. In this paper we show stochastic foundations of this process. We derive the equation by means of a modified Continuous Time Random Walk model. Interpretation of the $g$--subdiffusion process is also discussed.
\end{abstract}

\maketitle

{\it Introduction.}
Subdiffusion occurs in media in which the movement of diffusing molecules is very difficult due to the complex internal structure of the medium. Within the Continuous Time Random Walk (CTRW) model, a distribution of time between particle jumps $\psi$ has a heavy tail for subdiffusion, $\psi(t)\sim 1/t^{1+\alpha}$, $0<\alpha<1$ \cite{mk,ks,barkai2000}. This model leads to the ``ordinary'' subdiffusion equation with the fractional order Caputo derivative. Recently, a more general subdiffusion equation with the Caputo derivative with respect to another function $g$ has been considered \cite{kd}, see also Ref. \cite{sz}; we call it the $g$--subdiffusion equation which describes the $g$--subdiffusion process. As shown in Ref. \cite{kd}, this equation describes a process in which a type of diffusion can change over time. Unfortunately, $g$--subdiffusion has not had a stochastic interpretation yet. We show how to derive the $g$--subdiffusion equation by means of modified CTRW model and we discuss the interpretation of this process.

{\it ``Ordinary'' subdiffusion equation.}
The fractional subdiffusion equation with ``ordinary'' Caputo derivative of the order $\alpha\in(0,1)$ is \cite{kd}
\begin{equation}\label{eq1}
\frac{^C \partial^{\alpha} P(x,t)}{\partial t^\alpha}=D\frac{\partial^2 P(x,t)}{\partial x^2},
\end{equation}
where the Caputo fractional derivative is defined for $0<\alpha<1$ as 
\begin{equation}\label{eq2}
\frac{^Cd^{\alpha}}{dt^\alpha}f(t)=\frac{1}{\Gamma(1-\alpha)}\int_0^t (t-u)^{-\alpha}f'(u)du,
\end{equation}
$\alpha$ is a subdiffusion parameter and $D$ is a generalized diffusion coefficient measured in the units of ${\rm m^2/s^\alpha}$. 
To solve the equation the Laplace transform $\mathcal{L}$ can be used,
\begin{equation}\label{eq3}
\mathcal{L}[f(t)](s)=\int_0^\infty {\rm e}^{-st}f(t)dt.
\end{equation}
Due to the relation
\begin{equation}\label{eq4}
\mathcal{L}\left[\frac{^C d^\alpha f(t)}{dt^\alpha}\right](s)=s^\alpha\mathcal{L}[f(t)](s)-s^{\alpha-1}f(0),
\end{equation}
where $0<\alpha\leq 1$, we get
\begin{eqnarray}\label{eq5}
s^\alpha\mathcal{L}[P(x,t)](s)-s^{\alpha-1}P(x,0)\\
=D\frac{\partial^2\mathcal{L}[P(x,t)](s)}{\partial x^2}.\nonumber
\end{eqnarray}

{\it $G$--subdiffusion equation.}
In this paper functions describing $g$--subdiffusion are denoted by tilda. The $g$-subdiffusion equation reads 
\begin{equation}\label{eq6}
\frac{^C \partial^{\alpha}_g \tilde{P}(x,t)}{\partial t^\alpha}=D\frac{\partial^2 \tilde{P}(x,t)}{\partial x^2},
\end{equation}
where $0<\alpha<1$, the Caputo derivative with respect to another function $g$ is defined as \cite{abd}
\begin{equation}\label{eq7}
\frac{^Cd^{\alpha}_g f(t)}{dt^\alpha}=\frac{1}{\Gamma(1-\alpha)}\int_0^t (g(t)-g(u))^{-\alpha}f'(u)du,
\end{equation}
the function $g$ fulfils the conditions $g(0)=0$, $g(\infty)=\infty$, and $g'(t)>0$ for $t>0$, its values are given in a time unit. When $g(t)=t$, the $g$-Caputo fractional derivative takes a form of the ``ordinary'' Caputo derivative. To solve Eq. (\ref{eq6}) the $g$--Laplace transform can be used, this transform is defined as \cite{jarad}
\begin{equation}\label{eq8}
\mathcal{L}_g[\tilde{f}(t)](s)=\int_0^\infty {\rm e}^{-s g(t)}\tilde{f}(t)g'(t)dt.
\end{equation}
Due to the property \cite{jarad}
\begin{equation}\label{eq9}
\mathcal{L}_g\left[\frac{^Cd^{\alpha}_g}{dt^\alpha}\tilde{f}(t)\right](s)=s^\alpha\mathcal{L}_g[\tilde{f}(t)](s)-s^{\alpha-1}\tilde{f}(0),
\end{equation}
the procedure of solving Eq. (\ref{eq6}) is similar to the procedure of solving ``ordinary'' subdiffusion equation by means of the ``ordinary'' Laplace transform method.
In terms of the $g$-Laplace transform the $g$-subdiffusion equation is
\begin{eqnarray}\label{eq10}
s^\alpha\mathcal{L}_g[\tilde{P}(x,t)](s)-s^{\alpha-1}\tilde{P}(x,0)\\
=D\frac{\partial^2\mathcal{L}_g[P(x,t)](s)}{\partial x^2}.\nonumber
\end{eqnarray}
Using the $g$-Laplace transform to Eq. (\ref{eq6}) yields Eq. (\ref{eq10}) in the same form as Eq. (\ref{eq5}). 

{\it Model of particle random walk.} To derive the subdiffusion equation we use a simple model of a particle random walk along a one--dimensional homogeneous lattice. 
Usually, in the CTRW model both a particle jump length and waiting time for a particle jump are random variables. In our considerations, we assume that the jump length distribution $\lambda$ has the form $\lambda(x)=\frac{1}{2}[\delta(x-\epsilon)+\delta(x+\epsilon)]$, where $\delta$ is the delta Dirac function. Only the choice of a particle jump direction is random, its length $\epsilon$ is a parameter. We start with the particle random walk model in which the particle positions and time are discrete. Next, we move to continuous variables.
Random walk with discrete time $n$ is described by the equation $P_{n+1}(m)=\frac{1}{2}P_n(m+1)+\frac{1}{2}P_n(m-1)$, where $P_n(m)$ is a probability that a diffusing particle is at the position $m$ after $n$-th step. Let the initial particle position be $m=0$. Moving from discrete $m$ to continuous $x$ spatial variable we assume $x=m\epsilon$ and $P_n(x)=P_n(m)/\epsilon$, where $\epsilon$ is a distance between discrete sites. The above equations and the relation $[P_n(x+\epsilon)+P_n(x-\epsilon)-2P_n(x)]/\epsilon^2 =\partial^2 P_n(x)/\partial x^2$, $\epsilon\rightarrow 0$, provide the following equation in the limit of small $\epsilon$
\begin{equation}\label{eq11}
P_{n+1}(x)-P_n(x)=\epsilon^2\frac{\partial^2 P_n(x)}{\partial x^2}.
\end{equation}
To move from discrete to continuous time we use the formula \cite{mw}
\begin{equation}\label{eq12}
P(x,t)=\sum_{n=0}^\infty Q_n(t)P_n(x),
\end{equation}
where $Q_n(t)$ is the probability that a diffusing particle takes $n$ step in the time interval $(0,t)$. 
The function $Q_n$ is determined differently for the ``ordinary'' subdiffusion and $g$--subdiffusion. In the following, we find the rule for determining the functions $Q_n$ and the explicit form of the functions $\psi$ for both processes. These functions, together with Eqs. (\ref{eq11}) and (\ref{eq12}), provide ``ordinary'' subdiffusion and $g$--subdiffusion equations.

{\it The case of ``ordinary'' subdiffusion.}
In this case the function $Q_n$ is a convolution of $n$ distributions $\psi$ of a waiting time for a particle to jump and a function $U$ which is the probability that a particle does not change its position after $n$-th step,
\begin{equation}\label{eq13}
Q_n(t)=(\underbrace{\psi\ast\psi\ast\ldots\ast\psi}_{n\;times}\ast U)(t),
\end{equation}
where the convolution is defined as
\begin{equation}\label{eq14}
(f\ast h)(t)=\int_0^t f(u)h(t-u)du.
\end{equation} 
The ``ordinary'' Laplace transform has the following property that makes the transform useful in determining the function $Q_n$
\begin{equation}\label{eq15}
\mathcal{L}[(f\ast h)(t)](s)=\mathcal{L}[f(t)](s)\mathcal{L}[h(t)](s). 
\end{equation}
From Eqs.  (\ref{eq12}), (\ref{eq13}), and (\ref{eq15}) we have 
\begin{equation}\label{eq16}
\mathcal{L}[P(x,t)](s)=\mathcal{L}[U(t)](s)\sum_{n=0}^\infty\mathcal{L}^n [\psi(t)](s) P_n(x).
\end{equation}
Combining Eqs. (\ref{eq11}), (\ref{eq12}), and (\ref{eq16}) we get 
\begin{eqnarray}\label{eq17}
\frac{2(1-\mathcal{L}[\psi(t)](s))}{\epsilon^2 \mathcal{L}[\psi(t)](s)}\mathcal{L}[P(x,t)](s)\\
-\frac{2\mathcal{L}[U(t)](s)}{\epsilon^2 \mathcal{L}[\psi(t)](s)}P(x,0)=\frac{\partial^2 \mathcal{L}[P(x,t)](s)}{\partial x^2}.\nonumber
\end{eqnarray}
Eq. (\ref{eq17}) coincides with Eq. (\ref{eq5}) only if 
\begin{eqnarray*}
\frac{1-\mathcal{L}[\psi(t)](s)]}{\mathcal{L}[\psi(t)](s)}=\frac{\epsilon^2 s^\alpha}{2D},\;
\frac{\mathcal{L}[U(t)](s)}{\mathcal{L}[\psi(t)](s)}=\frac{\epsilon^2 s^{\alpha-1}}{2D}. 
\end{eqnarray*}
The solutions to the above equations are
\begin{equation}\label{eq18}
\mathcal{L}[\psi(t)](s)=\frac{1}{1+\frac{\epsilon^2 s^\alpha}{2D}},
\end{equation}
and
\begin{equation}\label{eq19}
\mathcal{L}[U(t)](s)=\frac{\epsilon^2 s^{\alpha-1}}{2D\left(1+\frac{\epsilon^2 s^\alpha}{2D}\right)}=\frac{1-\mathcal{L}[\psi(t)](s)}{s}.
\end{equation}
Due to the relations
\begin{equation}\label{eq20}
\mathcal{L}[1](s)=\frac{1}{s}\;,\;\mathcal{L}\left[\int_0^t f(u)du\right](s)=\frac{\mathcal{L}[f(t)](s)}{s},
\end{equation}
we get 
\begin{equation}\label{eq21}
U(t)=1-\int_0^t \psi(u)du.
\end{equation}
In order to find the function $\psi$ we use the relation
\begin{eqnarray}\label{eq22}
\mathcal{L}^{-1}[s^\nu {\rm e}^{-as^\beta}](t)=\frac{1}{t^{1+\nu}}\sum_{k=0}^\infty \frac{1}{k!\Gamma(-\nu-\beta k)}\left(-\frac{a}{t^\beta}\right)^k\\ \equiv f_{\nu,\beta}(t:a),\nonumber
\end{eqnarray}
where $a,\beta>0$, $\Gamma$ is the Gamma-Euler function. The function $f_{\nu,\beta}$ is the Wright function and the special case of the H-Fox function. To find the inverse Laplace transform of Eq. (\ref{eq18}) first we calculate the inverse Laplace transform of the function ${\rm e}^{-as^\beta}/(1+\tau s^\alpha)$, where $\tau=\epsilon^2/2D$ and $a,\beta>0$, using the formula $1/(1+u)=\sum_{n=0}^\infty u^n$ when $|u|<1$.  We get
\begin{eqnarray}\label{eq23}
\mathcal{L}\left[\frac{{\rm e}^{-as^\beta}}{1+\tau s^\alpha}\right](s)=\nonumber\\
\left\{
\begin{array}{c}
\frac{1}{\tau}\sum_{n=0}^\infty \left(-\frac{1}{\tau}\right)^n s^{-(n+1)\alpha}{\rm e}^{-as^\beta},\;s>\frac{1}{\tau^{1/\alpha}},\\
   \\
\sum_{n=0}^\infty (-\tau)^n s^{n\alpha}{\rm e}^{-as^\beta},\;s<\frac{1}{\tau^{1/\alpha}},
\end{array}
\right.
\end{eqnarray}
Next, we take the limit of $a\rightarrow 0$. From Eqs. (\ref{eq22}), (\ref{eq23}), and the relations  $f_{\nu,\beta}(t;0)=1/\Gamma(-\nu)t^{1+\nu}$, $1/\Gamma(0)=0$ we get
\begin{eqnarray}\label{eq24}
\psi(t)=\left\{
\begin{array}{c}
\frac{1}{\tau}\sum_{n=0}^\infty \left(-\frac{1}{\tau}\right)^n \frac{t^{(n+1)\alpha-1}}{\Gamma((n+1)\alpha)},\;t<\tau^{1/\alpha},\\
  \\
\sum_{n=0}^\infty \left(-\tau\right)^{n+1} \frac{t^{-(n+1)\alpha-1}}{\Gamma(-(n+1)\alpha)},\;t>\tau^{1/\alpha},
\end{array}
\right.
\end{eqnarray}
We have $\psi(t)\approx \alpha\tau/\Gamma(1-\alpha)t^{1+\alpha}$ in the limit of $t\rightarrow\infty$. The function $\psi$ was already derived using the relation $\mathcal{L}^{-1}[1/(1+\tau s^\alpha)]=t^{\alpha-1}E_{\alpha,\alpha}(-t^\alpha/\tau)$, where  $E_{\alpha,\alpha}(z)=\sum_{n=0}^\infty z^n/\Gamma(\alpha(n+1))$ is the two--parameter Mittag--Leffler function, see for example Ref. \cite{m}. Then, the function $\psi$ corresponds to Eq. (\ref{eq24}) but for the case of $t<\tau^{1/\alpha}$ only.

{\it The case of $g$--subdiffusion.}
To get Eq. (\ref{eq10}) we use the $g$--Laplace transform. This transform has the following property \cite{jarad}
\begin{equation}\label{eq25}
\mathcal{L}_g[(f\ast_g h)(t)](s)=\mathcal{L}_g[f(t)](s)\mathcal{L}_g[h(t)](s),
\end{equation}
where the $g$--convolution is defined as
\begin{equation}\label{eq26}
(f\ast_g h)(t)=\int_0^t f(u)h(g^{-1}(g(t)-g(u))g'(u)du.
\end{equation}
We involve the $g$--convolution in the CTRW model. Then, the procedure for deriving the $g$--subdiffusion equation using the $g$--Laplace transform is analogous to the procedure for deriving the "ordinary" subdiffusion equation using the ``ordinary'' Laplace transform. Assuming
\begin{equation}\label{eq27}
\tilde{P}(x,t)=\sum_{n=0}^\infty \tilde{Q}_n(t)P_n(x).
\end{equation}
and
\begin{equation}\label{eq28}
\tilde{Q}_n(t)=(\underbrace{\tilde{\psi}\ast_g\tilde{\psi}\ast_g\ldots\ast_g\tilde{\psi}}_{n\;times}\ast_g \tilde{U})(t),
\end{equation}
from Eqs. (\ref{eq25}), (\ref{eq27}), and (\ref{eq28}) we obtain
\begin{equation}\label{eq29}
\mathcal{L}_g[\tilde{P}(x,t)](s)=\sum_{n=0}^\infty\mathcal{L}_g[\tilde{U}(t)](s)\mathcal{L}^n_g[\tilde{\psi}(t)](s)P_n(x).
\end{equation}
From Eqs. (\ref{eq11}) and (\ref{eq29}) we get
\begin{eqnarray}\label{eq30}
\frac{1-\mathcal{L}_g[\tilde{\psi}(t)](s)}{\epsilon^2 \mathcal{L}_g[\tilde{\psi}(t)](s)}\mathcal{L}_g[\tilde{P}(x,t)](s)\\
-\frac{\mathcal{L}_g[\tilde{U}(t)](s)}{\epsilon^2 \mathcal{L}_g[\tilde{\psi}(t)](s)}\tilde{P}(x,0)=\frac{\partial^2\mathcal{L}[\tilde{P}(x,t)](s)}{\partial x^2}.\nonumber
\end{eqnarray}
Eq. (\ref{eq30}) is consistent with Eq. (\ref{eq10}) only when
\begin{equation}\label{eq31}
\mathcal{L}_g[\tilde{\psi}(t)](s)=\frac{1}{1+\frac{\epsilon^2 s^\alpha}{2D}}
\end{equation}
and
\begin{equation}\label{eq32}
\mathcal{L}_g[\tilde{U}(t)](s)=\frac{\epsilon^2 s^{\alpha-1}}{2D\left(1+\frac{\epsilon^2 s^\alpha}{2D}\right)}.
\end{equation}
Comparing Eqs. (\ref{eq31}) and (\ref{eq32}) with Eqs. (\ref{eq18}) and (\ref{eq19}), respectively, we get
\begin{equation}\label{eq33}
\mathcal{L}_g[\tilde{\psi}(t)](s)=\mathcal{L}[\psi(t)](s),
\end{equation}
\begin{equation}\label{eq34}
\mathcal{L}_g[\tilde{U}(t)](s)=\mathcal{L}[U(t)](s).
\end{equation}
From the relation 
\begin{equation}\label{eq35}
\mathcal{L}_g[\tilde{f}(t)](s)=\mathcal{L}[\tilde{f}(g^{-1}(t))](s),
\end{equation}
we get the following rule \cite{kd}
\begin{equation}\label{eq36}
\mathcal{L}_g[\tilde{f}(t)](s)=\mathcal{L}[f(t)](s)\Leftrightarrow \tilde{f}(t)=f(g(t)).
\end{equation}
Due to Eq. (\ref{eq36}), from Eqs. (\ref{eq33}) and (\ref{eq34}) we obtain
\begin{equation}\label{eq37}
\tilde{\psi}(t)=\psi(g(t)),
\end{equation}
and
\begin{equation}\label{eq38}
\tilde{U}(t)=U(g(t)).
\end{equation}
Eqs. (\ref{eq24}) and (\ref{eq37}) provide
\begin{eqnarray}\label{eq39}
\tilde{\psi}(t)=\left\{
\begin{array}{c}
\frac{1}{\tau}\sum_{n=0}^\infty \left(-\frac{1}{\tau}\right)^n \frac{g^{(n+1)\alpha-1}(t)}{\Gamma((n+1)\alpha)},\;t<g^{-1}(\tau^{1/\alpha}),\\
  \\
\sum_{n=0}^\infty \left(-\tau\right)^{n+1} \frac{g^{-(n+1)\alpha-1}(t)}{\Gamma(-(n+1)\alpha)},\;t>g^{-1}(\tau^{1/\alpha}),
\end{array}
\right.
\end{eqnarray}
We get $\psi(t)\approx \alpha\tau/\Gamma(1-\alpha)g^{1+\alpha}(t)$ when $t\rightarrow\infty$. 

We link the $g$--convolution with the ``ordinary'' convolution. Let $\tilde{f}(t)=f(g(t))$ and $\tilde{h}(t)=h(g(t))$. After simple calculation we get
\begin{equation}\label{eq40}
(\tilde{f}\ast_{g} \tilde{h})(t)=(f\ast h)(g(t)).
\end{equation}
From Eqs. (\ref{eq27}), (\ref{eq28}), and (\ref{eq40}) we have 
\begin{equation}\label{eq41}
\tilde{P}(x,t)=\sum_{n=0}^\infty Q_n(g(t))P_n(x).
\end{equation}
Comparing Eqs. (\ref{eq41}) and (\ref{eq12}) we obtain
\begin{equation}\label{eq42}
\tilde{P}(x,t)=P(x,g(t)).
\end{equation}

{\it Interpretation.} 
The $g$--subdiffusion process is associated to ``ordinary'' subdiffusion controlled by the same parameter $\alpha$. The waiting time for a particle jump in the $g$--subdiffusion process is controlled by the functions $\psi$ and $g$. A particle jump that would occur with some probability after time $t$ in "ordinary" subdiffusion process will occur with the same probability after time $\tilde{t}=g^{-1}(t)$ in the $g$--subdiffusion process. If $g(t)<t$ we have $t<\tilde{t}$, subdiffusion is then slowed down. When $g(t)>t$ subdiffusion is accelerated. An example of $g$--subdiffusion is the diffusion of molecules in a medium consisting of matrix in which there are narrow channels. If channels have a complicated geometric structure and diffusing molecules do not interact with the matrix, then "ordinary" subdiffusion controlled by the parameter $\alpha$ occurs. When temporary penetration of a molecule into the matrix is possible then the molecule "disappears" from channels and may diffuse further upon returning to a channel. In this case, "ordinary" subdiffusion is slowed down. Such a process occurs in a vessel filled with alginate beads immersed in water in which colistin antibiotic diffuses \cite{kdlwa}. When the matrix provides the diffusing molecules with additional energy, subdiffusion can be accelerated.

{\it Final remarks.}
We have shown that the $g$--subdiffusion equation can be derived by means of the modified CTRW model (we call it the $g$--CTRW model). In the $g$-CTRW model we use the $g$-convolution and the $g$-Laplace transform instead of the ``ordinary'' convolution and the``ordinary'' Laplace transform, respectively, which are used in the "ordinary" CTRW model.

We note that the condition $\mathcal{L}_g[\tilde{\psi}(t)](0)=1$ does not guarantee that the function $\tilde{\psi}$ is normalized. Therefore, $\tilde{\psi}$ is not a probability distribution. Thus, it seems that the $g$-CTRW model is merely a mathematical procedure. However, this model can be interpreted as an "ordinary" CTRW model in which the time scale is controlled by the function $g(t)$, see Eqs. (\ref{eq37})--(\ref{eq42}). The key issue for the $g$--subdiffusion process is determining the parameter $\alpha$ and the function $g$. An example of their determination from empirical data is shown in \cite{kdlwa}. 

In practice, the transformations made in deriving the $g$--subdiffusion equation within the $g$-CTRW model are the same as in deriving the "ordinary" subdiffusion equation using "ordinary" CTRW. Within the "ordinary" CTRW, subdiffusion-reaction equations \cite{kl} as well as the Green's functions and membrane boundary conditions for a system in which a thin membrane separates different subdiffusive media \cite{tk2019} have been derived. Within the $g$--CTRW model the same procedures can also be used to derive $g$--subdiffusion-reaction equations, Green's functions, and boundary conditions at the membrane for the processes described by $g$--subdiffusion equations. 

We suppose that the $g$--subdiffusion model can be used to describe diffusion of antibiotics in a biofilm. Biofilm usually has a gel structure. When the antibiotic does not interact with bacteria, the ``ordinary'' antibiotic subdiffusion in the biofilm is expected. However, bacteria in the biofilm have different defense mechanisms against the action of the antibiotic. These mechanisms may hinder or even facilitate antibiotic subdiffusion, see Ref. \cite{km} and the references cited therein. Thus, the application of the $g$--subdiffusion equation to describe this process may be effective.


\begin{thebibliography}{33}
\bibitem{mk} R. Metzler and J. Klafter, Phys. Rep. \textbf{339}, 1 (2000).
\bibitem{ks} J. Klafter and I.M. Sokolov, {\it First step in random walks. From tools to applications}, (Oxford UP, New York, 2011).
\bibitem{barkai2000} E. Barkai, R. Metzler, and J. Klafter, Phys. Rev. E \textbf{61}, 132 (2000).
\bibitem{kd} T. Koszto{\l}owicz and A. Dutkiewicz, Phys. Rev. E \textbf{104}, 014118 (2021).
\bibitem{sz} B. Samet and Y. Zhou, RACSAM \textbf{113}, 2887 (2019); 
R. Garra, A. Giusti, and F. Mainardi, Ricerche Mat. \textbf{67}, 899 (2018). 
\bibitem{abd} W. Abdelhedi, Comp. Appl. Math. \textbf{40}, 53 (2021); 
R. Almeida, Commun. Nonlinear Sci. Numer. Simul. \textbf{44}, 460 (2017); 
A.A. Kilbas, H.M. Srivastava, and J.J. Trujillo, {\it Theory and Applications of Fractional Differential Equations} (North-Holland Mathematics Studies, 204, Elsevier, Amsterdam, 2006); 
J.V.D.C. Sousa and E.C. de Oliveira, Commun. Nonlinear Sci. Numer. Simul. \textbf{60}, 72 (2018).
\bibitem{jarad} F. Jarad and T. Abdeljawad, Discrete And Continuous Dynamical Systems-S \textbf{13}, 709 (2020); 
F. Jarad, T. Abdeljawad, S. Rashid, and Z. Hammouch, Adv. Differ. Equ. \textbf{2020}, 303 (2020).
\bibitem{mw} E.W. Montroll and G.H. Weiss, J. Math. Phys. \textbf{6}, 167 (1965).
\bibitem{tk2004} T. Koszto{\l}owicz, J. Phys. A \textbf{37}, 10779 (2004).
\bibitem{m} F. Mainardi, WSEAS Trans. Math. \textbf{19}, 74 (2020);
Entropy \textbf{22}, 1359 (2020).
\bibitem{kdlwa} T. Koszto{\l}owicz, A. Dutkiewicz, K. Lewandowska, S. W\c{a}sik, and M. Arabski, arXiv: cond-mat. 2107.02419 (2021).
\bibitem{kl} T. Koszto{\l}owicz and K. Lewandowska, Phys. Rev. E \textbf{90}, 032136 (2014). 
\bibitem{tk2019} T. Koszto{\l}owicz, Phys. Rev. E \textbf{99}, 022127 (2019); 
Int. J. Heat Mass Transf. \textbf{111}, 1322 (2017).
\bibitem{km} T. Koszto{\l}owicz and R. Metzler, Phys. Rev. E \textbf{102}, 032408 (2020);
T. Koszto{\l}owicz, R. Metzler, S. W\c{a}sik, and M. Arabski, PLoS One \textbf{15}, e0243003 (2020).


\end{thebibliography}
\end{document}